\documentclass[notitlepage,nofootinbib,twocolumn,amsmath]{revtex4-1}

\usepackage{amsfonts}
\usepackage{amsmath}
\usepackage{amssymb}
\usepackage{aas_macros}
\usepackage{graphicx}
\usepackage{color}
\usepackage{float}
\allowdisplaybreaks[1]

\def\lsim{~\rlap{$<$}{\lower 1.0ex\hbox{$\sim$}}}
\def\bsim{~\rlap{$>$}{\lower 1.0ex\hbox{$\sim$}}}

\def\hkpc{\ {\rm {\it h}^{-1}Kpc}}

\def\eV{\ {\rm eV}}
\def\GeV{\ {\rm GeV}}

\def\apjl{Astrophys.\ J.\ Lett.}

\def\aap{Astron.\ Astrophys.}

\def\apjs{Astrophys.\ J. Supp.}

\def\la{\langle}
\def\ra{\rangle}

\def\mathbi#1{\textbf{\em #1}}

\def\vu{\mathbi{u}}

\def\vx{\mathbi{x}}

\def\grad{\boldsymbol{\nabla}}

\definecolor{RedWine}{rgb}{0.743,0,0}
\definecolor{RoyalBlue}{rgb}{0.25,.41,.88}
\definecolor{ForestGreen}{rgb}{.13,.54,.13}
\definecolor{DeepPurple}{rgb}{.72,.18,1}

\begin{document}

\title[Backreaction of axion coherent oscillations]{Backreaction of axion coherent oscillations}
	

\author{Mor Rozner}
\email{morozner@campus.technion.ac.il}
\affiliation{Physics department and Asher Space Science Institute, Technion, Haifa 3200003, Israel}
\author{Vincent Desjacques}
\email{dvince@physics.technion.ac.il}
\affiliation{Physics department and Asher Space Science Institute, Technion, Haifa 3200003, Israel}
	
\date{\today}
\label{firstpage}
	
\begin{abstract}

  We investigate how coherent oscillations backreact on the evolution of the condensate wave function
  of ultra-light axions in the non-relativistic regime appropriate to cosmic structure formation.
  The coherent oscillations induce higher harmonics beyond the fundamental mode considered so far
  when a self-interaction is present, and imprint oscillations in the gravitational potential.
  We emphasize that the effective self-interaction felt by the slowly-varying envelop of the wave
  function always differs from the bare Lagrangian interaction potential.
  We also point out that, in the hydrodynamical formulation of the Gross-Pitaevskii equation,
  oscillations in the gravitational potential result in an attractive force that counteracts the
  effect of the quantum pressure arising from the strong delocalization of the particles.
  Since these effects become significant on physical scales less than the (large) Compton wavelength of
  the particle, they are presumably not very relevant on the mildly nonlinear scales traced by
  intergalactic neutral hydrogen for axion masses consistent with the bounds from the Lyman-$\alpha$
  forest. However, they might affect the formation of virialized cosmological structures and their
  stability.

\end{abstract}


\maketitle
	
\section{Introduction}

Since the virial equilibrium considerations of Zwicky and his inference of a ``missing mass'' in galaxy clusters \cite{Zwicky:1933},
dark matter has become a inescapable ingredient of nearly all viable cosmological models \cite{PlanckCollaboration;2016}.
Nonetheless, its nature has remained thus far elusive, and there is a plethora of models trying to explain it.	
Beyond the popular weakly interacting massive particles (WIMPs) \cite{Steigman/Turner;1985}, primordial black holes
\cite{Chapline/Frampton:2016} or a modification to gravity such as MOND \cite{Milgrom;1983} could also explain at least part
of the observations.

In another class of models, dark matter is made of light bosons such as axions
\cite{Abbott:1982af,Preskill:1982cy,Dine:1981rt,Baldeschi:1983mq,Sin:1992bg,HBG:2000,Riotto:2000kh,Amendola/Barbieri:2006,Sikivie/Yank:2009,Kim/Marsh:2016}.
Such particles could be produced in the early Universe from a symmetry-breaking event conjectured to solve the strong CP problem
\cite{Weinberg:1977ma,Wilczek:1977pj,PQ1,PQ2}. From a cosmological point of view, ultra-light axions with a mass $10^{-22}\eV$ 
much smaller than the mass of QCD axions are particularly interesting because they could solve some of the small-scale
problems associated with standard cold dark matter (CDM) like cusps vs. cores inside dark matter halos
(see \cite{Hui/Ostriker/etal:2017,marsh:2016} and references therein).
However, observations of the Lyman-$\alpha$ forest appear to rule out ultra-light axions as dark matter if their mass is
less than $10^{-21}\eV$ \cite{Irsic:2017yje,Armengaud:2017nkf}, although uncertainties in the thermal history of the intergalactic
medium somewhat weaken this constraint \cite{Zhang:2017chj}. 

``Fuzzy'' dark matter (FDM) such as ultra-light axions is in the form of a Bose-Einstein condensate due to the very large de
Broglie wavelength of the particles (the details of the condensation process are still uncertain, see for instance
\cite{Kolb:1993hw,Semikoz:1995rd,Sikivie/Yang:2009,Davidson:2013aba,Guth:2014hsa,Levkov:2018kau}).
In the non-relativistic limit, the (scalar) wave function of the mean field evolves
according to the Gross-Pitaevskii (GP) equation. On applying the Madelung transformation \cite{Madelung:1927,Bohm:1952}, this
equation can be re-expressed as standard hydrodynamical equations, in which gravity and axion self-interactions are
balanced by the ``quantum'' pressure. This hydrodynamical representation can be easily implemented numerically to study the
formation of cosmic structures in FDM models
\cite{woo/chiueh:2008,Schive/etal:2014,veltmaat/niemeyer:2016,Zhang:2016uiy,Mocz:2017wlg,schive/chiueh:2018}, for which existing
Boltzmann code can be modified to provide the suitable initial conditions \cite[e.g.,][]{Hlozek/Grin/etal:2015}.

While the GP equation encodes the leading contributions to the hydrodynamic gradient expansion, higher-order effects, which are
generally neglected, might play an important role at smaller scales
(i.e., in the formation of ``Bose stars'' \cite{Ruffini:1969qy,Tkachev:1986tr}.) and possibly affect the cosmic structure formation
in FDM models. In particular, axion coherent oscillations will generate higher harmonics beyond the fundamental mode considered so far
when a self-interaction is present \cite[see e.g.][for a recent study in the context of ``dense'' Bose stars]{visinelli/etal:2018,Eby:2017teq},
and imprint oscillations in the gravitational potential \cite{Khmelnitsky/Rubakov:2013}.
The goal of this paper is to investigate how the coherent oscillations backreact on the evolution of the condensate wave function --
in the non-relativistic regime appropriate to the large scale structure of the Universe -- and how this backreaction manifests itself
in the hydrodynamical formulation of the GP equation. Obviously, since the GP equation captures the leading contributions at large
scales, this backreaction can only be significant on scales comparable to, or smaller than, the (huge) Compton wavelength of the particle.

Our paper is organized as follows. After a review of the ultra-light axion model of cold dark matter (Sec.~\S\ref{sec:axions}),
we pursue with a discussion of the backreaction of axion coherent oscillations at the homogeneous level (Sec.~\S\ref{sec:homogeneous}),
before taking into account some of the backreaction effects in the derivation of the Gross-Pitaevskii system
(Sec.~\S\ref{sec:perturbations}).
We discuss the implications of our findings in Sec.~\S\ref{sec:discussion} and conclude in Sec.~\S\ref{sec:conclusions}.
	
We shall use the natural units $c=\hbar=k_B=1$ throughout,
with a gravitational coupling constant $G=1/m_P^2$ given in terms of the Planck mass $m_P=1.22\times 10^{19}\GeV$.
In these units, the Hubble constant is $H_0=2.13h\times 10^{-41}\GeV$.
Finally, Greek indices $\mu$, $\nu$ etc. run over the 4 spacetime dimensions, whereas Latin indices $i$, $j$ etc. run over the spatial
dimensions solely.
	
\section{Axion as a cold dark matter candidate}
\label{sec:axions}

We begin with a brief overview of the theoretical results relevant to the description of axions in the large scale structure of the Universe
\cite[see][for a review]{marsh:2016}.

\subsection{Lagrangian and characteristic scales}

The starting point is the action for the axion (real scalar) field $\phi$,
\begin{equation}
  S[\phi]=\int\! d^4x\,\sqrt{-g}\left[\frac{1}{2}(\partial\phi)^2-\Lambda^4\left(1-\cos\frac{\phi}{f}\right)\right]\;,
\end{equation}
where $\Lambda$ is a sort of condensation scale, $f$ is the decay constant and $g$ is the metric determinant.

Expanding for $\phi < f$ and including the quartic coupling, one obtains a potential of the form
\begin{gather}
  \label{eq:Vaxion}
  V(\phi) = \frac{1}{2} m_a^2 \phi^2 - \frac{1}{4!} \lambda \phi^4 \;, \\
  \mbox{where} \quad  m_a^2=\frac{\Lambda^4}{f^2} \quad \mbox{and} \quad\lambda = \frac{m_a^2}{f^2} \;. \nonumber
\end{gather}
Our sign convention is such that $\lambda>0$ when the quartic interaction is attractive.
While our analysis is valid for any light boson in the form of a condensate, we shall adopt a fiducial axion mass $m_a=10^{-22}\eV$ and decay
constant $f=10^{17}\GeV$ in all illustrations.
This gives $\lambda=+10^{-96}$.
Higher order terms ($\phi^6$ and higher) are negligible as long as $\phi/f\ll 1$ and this remains true as well when taking into account the high
phase-space density.
Notice that $\phi$ has dimension of energy, and that the sign of the self-interaction coupling leads to an attractive force.
This will be relevant for all our considerations.

The relative importance of time-derivatives and gradients of the field $\phi$ can be estimated from a few characteristics ratios.
Firstly, $H/m_a$, where $H$ is the Hubble rate, represents the importance of Hubble friction relative to the coherent oscillations
of the condensate. In matter domination, we have
\begin{equation}
  \frac{H_0}{m_a}\simeq 2.1 \times 10^{-11} h a^{-3/2}\left(\frac{m_a}{10^{-22}\eV}\right)^{-1} \;.
\end{equation}
We have $H/m_a\ll 1$ long after at the onset of oscillations, which occurs at a scale factor $a_\text{osc}\sim 10^{-6}-10^{-7}$ for
the fiducial parameters.
Secondly, $k/m_a a$, where $k$ is the comoving wavenumber and $a$ is the scale factor normalized to unity at the present epoch,
characterizes the importance of kinetic terms. We have
\begin{equation}
  \frac{k}{m_a a} \simeq 6.4\times 10^{-5} a^{-1}\left(\frac{k}{{\rm Kpc}^{-1}}\right)\left(\frac{m_a}{10^{-22}\eV}\right)^{-1}\;.
\end{equation}
This ratio involves the (reduced) {\it Compton} wavelength $m_a^{-1}$ of the particle rather than its {\it de Broglie} wavelength
$(m_a v)^{-1}$, where $v\sim 10^{-4}$ is a typical velocity.
After matter-radiation equality, $k/m_a a$ never reaches unity on scales $k\lesssim 1\,{\rm Kpc}^{-1}$ unless the particle mass is much less
than $10^{-22}\eV$.
Finally, the dimensionless coupling constant $\lambda$ quantifies the importance of the self-interaction.
Note that $\lambda$ always appears multiplied by a factor of $\rho_a/m_a^4$, where $\rho_a$ is the axion energy density.

Since our focus is on the late-time, sub-horizon evolution of the axion field relevant to the formation of cosmic structures, we will
consider a regime in which $H/m_a\ll 1$ and $H\lesssim k/a \lesssim m_a$. Furthermore, since we are interested in
scales at which the density does not considerably exceed its background value, we also have $\lambda\rho_a/m_a^4\ll 1$.
As a result, powers of $H/m_a$, $k/m_a a$ and $\lambda\rho_a/m_a^4$ are small and become rapidly negligible in a perturbative expansion
\cite[see, e.g.,][]{ratra:1991,Hwang/Noh:2009}.

\subsection{The Gross-Pitaevskii equation}

The classical Euler-Lagrange equation leads to the Klein-Gordon (KG) equation
\begin{equation}
  \Box\phi\equiv g^{\mu\nu}\nabla_\mu(\partial_\nu\phi)=\partial_\phi V \;,
\end{equation}
where $\Box$ is the d'Alembertian operator.
We choose the conformal Newtonian gauge, in which the metric of the perturbed flat FRW spacetime takes the form
\begin{equation}
  ds^2 = a^2\big[-(1+2\Psi)d\eta^2 + (1-2\Phi)\delta_{ij}dx^i dx^j\big]
\end{equation}
and the Bardeen potentials are of order $|\Psi|\sim|\Phi|\sim 10^{-5}$ on the scales $k\lesssim m_a a$ we are interested in.

Considering the weak field limit, where $\delta g_{\mu \nu}$ is small w.r.t. $g_{\mu \nu}$, the KG equation reduces to
\begin{gather}
  \big(1-2\Psi\big)\big(\ddot{\phi}+2{\cal H}\dot{\phi}\big) - \big(1-2\Phi\big) \partial_i\partial^i\phi + a^2 \partial_\phi V \\
  = \big(\dot{\Psi}+3\dot{\Phi}\big)\dot{\phi} + \partial^i\big(\Psi-\Phi\big)\partial_i\phi \nonumber \;.
\end{gather}
Here, ${\cal H}=aH$ is the conformal Hubble rate and a dot designates a derivative w.r.t. the conformal time $\eta$.
This equation simplifies drastically in the absence of any anisotropic stress. In our regime of interest, this is justified because,
in the non-relativistic limit, the (second-order) anisotropic stress of the axion field (also known as ``quantum pressure'') is 
\begin{equation}
  \pi_{ij}= \frac{1}{4 m_a^2 a^2\rho_a} \Big(\partial_i\rho_a\partial_j\rho_a-\frac{1}{3}\delta_{ij}(\partial_k\rho_a)^2\Big) \;,
\end{equation}
where $\rho_a\simeq m_a^2 \phi^2$. Therefore, 
\begin{equation}
  \big\lvert \pi_{ij}\big\lvert \sim \frac{k^2}{m_a^2 a^2} \rho_a \;.
\end{equation}
Einstein equations then imply 
\begin{equation}
  \big\lvert \Phi-\Psi\big\lvert \sim a^2 \frac{G}{k^2}\big\lvert\pi_{ij}\big\lvert \sim \frac{\rho_a}{(m_a m_P)^2} \;.
\end{equation}
For an axion energy density equal to the present-day critical density, $\rho_a =\rho_\text{cr,0}\sim 10^{-46} \GeV$, we find that
$|\Phi-\Psi|\sim 10^{-24}$ is negligible. This shows that the impact of the axion anisotropic stress on the Bardeen
potentials can be safely neglected in the regime under consideration.
Note that, in a realistic cosmological model including other particles such as massive neutrinos, there would be a (first order)
anisotropic stress. With $\Psi=\Phi$ (which we shall refer to as the gravitational potential), the KG equation simplifies to
\begin{equation}
  \label{eq:KGsimplified}
  \ddot{\phi}+2{\cal H}\dot{\phi}-4\dot{\Psi}\dot{\phi}+a^2(1+2\Psi\big)\partial_\phi V = 0 \;.
\end{equation}
We retained the term $-4\dot{\Psi}\dot{\phi}$ for the following reason:
when oscillations in the gravitational potential are taken into account, it is only
suppressed by a factor of $(k/m_a a)^2$ rather than $(H/m_a)^2$ relative to leading-order contribution $\Psi\phi$.
Conversely, we ignored the term proportional to $\Psi\partial_i\partial^i\phi$ because it contributes only at second order.

The Gross-Pitaevskii equation is derived under the assumption that the axion field undergoes coherent oscillations. Therefore,
one makes the ansatz
\begin{align}
  \label{eq:phiansatzGP}
  \phi(\eta,\vx) &= \sqrt{2}\,{\rm Re}\!\Big[\psi(\eta,\vx)\, e^{-im_at}\Big] \\
  t &= \int_0^\eta\!\!d\eta' a(\eta') \nonumber \;,
\end{align}
where the slowly-varying, complex envelop $\psi(\eta,\vx)$ changes on a timescale of order ${\cal H}^{-1}$.
The frequency $m_a$ appears because we consider the non-relativistic limit, in which the typical energy of a particle is $E\approx m_a$.
Further assuming that the gravitational potential varies slowly (so that $\dot{\Psi}\dot{\phi}$ can be neglected), one eventually obtains
\begin{align}
  i a\bigg(\partial_\eta\psi+\frac{3}{2}\mathcal{H}\psi\bigg)
  &= -\frac{1}{2m_a} \partial_i\partial^i\psi \\
  &\quad + m_a a^2\left(\Psi - \frac{1}{8f^2}|\psi|^2\right)\psi \nonumber \\
  & \quad + \bigg. \mbox{fast oscillating piece} \nonumber \;.
\end{align}
Factors of $\hbar$ arise with the partial derivatives, so that this equation is truly quantum
\cite[See for instance][for a discussion of the ``classical'' limit $\hbar/m_a\to 0$]{suarez/chavanis:2015,mocz/lancaster/etal:2018}.
Furthermore, the gravitational energy can now be recast in the form of an interaction potential $m_a^2\Psi$ appropriate to the Newtonian limit.
Finally, the fast oscillatory piece involves the harmonics $e^{\pm 2 im_at}$, $e^{\pm 4im_at}$ and varies on a timescale much shorter than $\psi(\eta,\vx)$.
Therefore, it is usually neglected and one is left with a nonlinear Schr\"odinger equation.
Note that the ansatz Eq.(\ref{eq:phiansatzGP}) properly describes the long-term behaviour of the axion oscillation only if the axion self-interaction
can be neglected. 

We will now investigate the backreaction of the self-interaction and of the rapidly varying gravitational potential on the equation of motion, which
arises when the axion field undergoes coherent oscillations.

\section{Coherent oscillations of the homogeneous background}
\label{sec:homogeneous}
	
We begin with a discussion of the homogeneous KG equation.
We shall see that ${\cal H}$ also exhibits oscillations, which are largest at the onset of the axion coherent oscillations.
However, for a realistic cosmological model including a component of relativistic particles, this is a small effect for our
fiducial axion mass. This calculation will also provide us with an ingredient useful to the discussion of the inhomogeneous case,
that is, the value of classical action for the background solution.

\subsection{Free oscillations}
\label{sec:freeoscillations}

We begin with the homogeneous KG equation
\begin{equation}
  \label{eq:KGhom}
  \ddot{\bar\phi}+2{\cal H} \dot{\bar\phi}+m_a^2 a^2\bar\phi = 0\;.
\end{equation}
Following \cite{ratra:1991,Hwang/Noh:2009}, one substitutes a solution of the form
\begin{equation}
  \label{eq:ansatz1}
  \bar\phi(\eta) \sim \sqrt{2} \Big[\phi_\text{1c}(\eta)\cos(m_at)+\phi_\text{1s}(\eta)\sin(m_at)\Big] \;,
\end{equation}
where $t\equiv t(\eta)$ is the cosmic time.
The assumption is that both $\phi_\text{1c}(\eta)$ and $\phi_\text{1s}(\eta)$ vary on the long timescale ${\cal H}^{-1}$.
The dependence of the sine and cosine on $m_a t$ (rather than $m_a\eta$) enable use to easily handle the mass term $m_a^2 a^2\bar\phi$.
At order $H/m_a$ (which quickly drops below unity after the onset of axion oscillations), both satisfy the differential equation
\cite{Hwang/Noh:2009}
\begin{equation}
  \label{eq:evol1}
  \dot{\phi}+\frac{3}{2}{\cal H}\phi=0 \;.
\end{equation}
The axion background energy density and pressure are given by
\begin{align}
  \bar\rho_a &= \bar\rho_0 + \bar\rho_\text{2c}\cos(2m_at) + \bar\rho_\text{2s} \sin(2m_at) \\
  \bar P_a &= \bar P_0 +\bar P_\text{2c} \cos(2m_at) + \bar P_\text{2s}\sin(2m_at) \nonumber \;,
\end{align}
with
\begin{align}
  \label{eq:barrho}
  \bar\rho_0 &= m_a^2\bigg[\Big(\phi_\text{1c}^2+\phi_\text{1s}^2\Big)
    +\frac{1}{m_aa}\Big(\dot{\phi}_\text{1c}\phi_\text{1s}-\dot{\phi}_\text{1s}\phi_\text{1c}\Big)\bigg] \\
  \bar\rho_\text{2c} &= m_a^2\bigg[\frac{1}{m_aa}\Big(\dot{\phi}_\text{1c}\phi_\text{1s}+\dot{\phi}_\text{1s}\phi_\text{1c}\Big)\bigg]\nonumber \\
  \bar\rho_\text{2s} &= m_a^2\bigg[\frac{1}{m_aa}\Big(\dot{\phi}_\text{1s}\phi_\text{1s}-\dot{\phi}_\text{1c}\phi_\text{1c}\Big)\bigg]\nonumber \;,
\end{align}        
and
\begin{align}
  \label{eq:barP}
  \bar P_0 &= m_a^2 \bigg[\frac{1}{m_aa}\Big(\dot{\phi}_\text{1c}\phi_\text{1s}-\dot{\phi}_\text{1s}\phi_\text{1c}\Big)\bigg] \\
  \bar P_\text{2c} &= m_a^2 \bigg[\Big(\phi_\text{1s}^2-\phi_\text{1c}^2\Big)+\frac{1}{m_aa}\Big(\dot{\phi}_\text{1c}\phi_\text{1s}
    +\dot{\phi}_\text{1s}\phi_\text{1c}\Big)\bigg] \nonumber \\
  \bar P_\text{2s} &= m_a^2 \bigg[-2 \phi_\text{1c}\phi_\text{1s}
    +\frac{1}{m_aa}\Big(\dot{\phi}_\text{1s}\phi_\text{1s}-\dot{\phi}_\text{1c}\phi_\text{1c}\Big)\bigg]\nonumber 
\end{align}      
up to order $H/m_a$.

Although this suggests that $\bar P_0\sim (H/m_a)\bar\rho_0$, a closer look at the Wronskian reveals that this is not
the case. Namely, let $y_1=\phi_\text{1c}\cos(m_at)$ and $y_2=\phi_\text{1s}\sin(m_at)$ be the two independent solutions of the homogeneous
equation. The Wronskian is given by
\begin{align}
  W(y_1,y_2) &= \left\lvert\begin{array}{cc} y_1 & y_2 \\ \dot{y}_1 & \dot{y}_2\end{array}\right\lvert  \\
  &= m_aa\phi_\text{1c}\phi_\text{1s} +
  \Big(\dot{\phi}_\text{1s}\phi_\text{1c} - \dot{\phi}_\text{1c}\phi_\text{1s}\Big) \nonumber \\
  &\qquad \times \cos(m_at)\sin(m_at) \nonumber \;.
\end{align}
From a famous theorem of Abell, it must satisfy
\begin{equation}
  W(y_1,y_2) = \tilde c\, e^{-2\int\!{\cal H}d\eta} = \frac{\tilde c}{a^2} \;,
\end{equation}
where $\tilde c$ does {\it not} depend on time for $y_1$ and $y_2$ to be linearly independent.
Since $a\propto \eta$ and ${\cal H}= \eta^{-1}\propto a^{-1}$ in radiation domination (RD era), while $a\propto\eta^2$ and
${\cal H}=2\eta^{-1}\propto a^{-1/2}$ in matter domination (MD era), the solution to Eq.(\ref{eq:evol1}) is
(e.g., \cite{HBG:2000,Hwang/Noh:2009,Turner:1983he})
\begin{equation}
  \phi \propto a^{-3/2} 
\end{equation}
in both eras. This implies
\begin{equation}
  \label{eq:fromwronskian}
  \phi_\text{1s}(\eta)=\mbox{const}\,\cdot\,\phi_\text{1c}(\eta) + {\cal O}\!\left(\frac{H}{m_a}\right) \;.
\end{equation}
Explicit solutions for $\phi_\text{1c}(t)$ and $\phi_\text{1s}(t)$ are given in Appendix \S\ref{app:homKG}.

Eq.~(\ref{eq:fromwronskian}) leads to the cancellation of the term $\dot{\phi}_\text{1c}\phi_\text{1s}-\dot{\phi}_\text{1s}\phi_\text{1c}$ at order
$H/m_a$ in $\bar\rho_0$ and $\bar P_0$. Therefore, the slowly varying part of the pressure is only of order $\bar P_0\sim (H/m_a)^2\bar\rho_0$.
By contrast, the amplitude $\bar P_\text{2c}$ and $\bar P_\text{2s}$ of the second harmonics is of order $\bar\rho_0$ as was first pointed out
by \cite{Khmelnitsky/Rubakov:2013}. We will discuss the implications of this in Sec.~\S\ref{sec:discussion}.

\subsection{Hubble rate}

We turn to the evolution of the background FRW universe and assume that the latter is filled by the axions and by a relativistic component
with homogeneous density $\bar\rho_r$ and pressure $\bar P_r=(1/3)\bar\rho_r$. Hence, the Friedmann equations read
\begin{align}
  {\cal H}^2 &= \frac{8\pi G}{3} \Big(\bar\rho_a+\bar\rho_r\Big) a^2 \\
  \dot{\cal H} &= -\frac{4\pi G}{3}\Big(\bar\rho_a+\bar P_a+2\bar\rho_r\Big) a^2 \nonumber\;. 
\end{align}
In light of the oscillatory pieces in $\bar\rho_a$ and $\bar P_a$, ${\cal H}$ must be of the form
\begin{equation}
  {\cal H} = {\cal H}_0 + {\cal H}_\text{2c}\cos(2m_at)+{\cal H}_\text{2s}\sin(2m_at) + \dots 
\end{equation}
where ${\cal H}_0$ (to be distinguished from the present-day Hubble rate $H_0$) is the dominant, slowly-varying contribution, while ${\cal H}_\text{2c}$
and ${\cal H}_\text{2s}$ both vary on a timescale ${\cal H}_0^{-1}$.
Using Eq.(\ref{eq:evol1}), this yields the following equations for the fast oscillating contributions to the Hubble rate,
\begin{multline}
  \dot{\cal H}_\text{2c} + 2 m_aa {\cal H}_\text{2s} \\ = -4\pi G m_a^2 a^2
  \Big(\phi_\text{1s}^2-\phi_\text{1c}^2\Big)\bigg[1+{\cal O}\!\left(\frac{H}{m_a}\right)\bigg]
\end{multline}
and
\begin{multline}
  \dot{\cal H}_\text{2s} - 2 m_aa {\cal H}_\text{2c} \\ = 8\pi G m_a^2 a^2 \phi_\text{1c}\phi_\text{1s}
  \bigg[1+{\cal O}\!\left(\frac{H}{m_a}\right)\bigg]
\end{multline}
In the left-hand side, the term ${\cal H}_\text{2c}$ (resp. ${\cal H}_\text{2s}$) is of order $H/m_a$ relative to
$2m_aa{\cal H}_\text{2s}$ (resp. $2m_aa{\cal H}_\text{2c}$). Hence, the leading order contribution to ${\cal H}_\text{2s}$ and ${\cal H}_\text{2c}$
are given by
\begin{align}
  {\cal H}_\text{2s} &\approx - 2 \pi G m_a a \Big(\phi_\text{1s}^2-\phi_\text{1c}^2\Big) \\
  {\cal H}_\text{2c} &\approx -4\pi G m_a a \phi_\text{1c}\phi_\text{1s} \nonumber \;.
\end{align}
Using the first Friedmann equation and setting ${\cal H}_0/(m_a a) \equiv H/m_a$, this implies
\begin{align}
  {\cal H}_\text{2s} &\sim -\frac{3}{4} \left(\frac{H}{m_a}\right)
  \frac{m_a^2\Big(\phi_\text{1s}^2-\phi_\text{1c}^2\Big)}{\bar\rho_0+\bar\rho_r}{\cal H}_0 \\
       {\cal H}_\text{2c} &\sim -\frac{3}{2}\left(\frac{H}{m_a}\right)
       \frac{m_a^2\phi_\text{1c}\phi_\text{1s}}{\bar\rho_0+\bar\rho_r}{\cal H}_0 \nonumber \;.
\end{align}
Since $\phi_\text{1c}=\mbox{const}\cdot\phi_\text{1s}$ at leading order (see Sec.~\S\ref{sec:freeoscillations}), we arrive at
\begin{align}
  \label{eq:H2s}
        {\cal H}_\text{2s} &= \frac{3}{4}\left(\frac{H}{m_a}\right)\,
        \frac{\cos (2\vartheta)}{1+\frac{\Omega_r}{\Omega_a}\left(\frac{a_0}{a}\right)}\,{\cal H}_0 \\
        \label{eq:H2c}
              {\cal H}_\text{2c} &= -\frac{3}{4}\left(\frac{H}{m_a}\right)\,
              \frac{\sin (2\vartheta)}{1+\frac{\Omega_r}{\Omega_a}\left(\frac{a_0}{a}\right)}\,{\cal H}_0
              \nonumber \;.
\end{align}
In particular, $\vartheta=3\pi/8$ in the simple adiabatic evolution outlined in Appendix \S\ref{app:homKG}.
We will see in Sec.~\S\ref{sec:discussion} that $\vartheta$ is, in fact, equal to the value of the (real) classical action for the homogeneous
solution.

In radiation domination, for which $\bar\rho_r\gg\bar\rho_0$, ${\cal H}_\text{2s}$ and ${\cal H}_\text{2c}$ are suppressed by a factor of
$(H/m_a)(\bar\rho_0/\bar\rho_r)\sim a^{-1}$ relative to ${\cal H}_0$.
In matter domination, this suppression relative to ${\cal H}_0$ scales as $H/m_a\sim a^{-3/2}$.
Therefore, axion oscillations can have a significant impact on the expansion rate only if their onset occurs around the matter-radiation
equality. This would happen for an axion mass $m_a\lesssim \frac{\Omega_a^2}{\Omega_r^{3/2}} H_0 \sim 10^{-27}\eV$.
However, this range of axion mass is, as of today, ruled out by Lyman-$\alpha$ forest measurements if all the dark matter is in the form
of axions \cite{Irsic:2017yje}.
Therefore, viable axion models (consistent with Lyman-$\alpha$ forest data) have a negligible impact on the expansion rate.
Hence, ${\cal H}_\text{2s}$ and ${\cal H}_\text{2c}$ can be safely ignored,
except when one also considers fluctuations in the gravitational potential (see Sec.~\S\ref{sec:gravosc}).
        
\subsection{Including a quartic interaction}
	
While the background KG equation of a free scalar field involves only the fundamental mode $e^{\pm im_at}$ in the long-time asymptotics
regime considered here, interactions will generate higher-order harmonics as is well-known from anharmonic oscillators.
Namely, when a quartic self-interaction is included, the KG equation for the background scalar field $\bar\phi$ takes the textbook form of
a driven, damped harmonic oscillator:
\begin{equation}
  \label{eq:KGback}
  \ddot{\bar\phi}+2{\cal H}\dot{\bar\phi}+m_a^2 a^2 \bar\phi = a^2 \frac{\lambda}{3!}\bar\phi^3 \;.
\end{equation}
As is well known, substituting a solution of the form $\bar\phi\sim \phi_\text{1c}(\eta)\cos(m_a t) + \phi_\text{1s}(\eta)\sin(m_at)$ into this
equation generates contributions proportional to $e^{\pm 3im_a t}$ owing to the $\bar\phi^3$ term in the right-hand side.
Therefore, one should consider instead
\begin{equation}
  \label{eq:phiansatz0}
  \bar\phi(\eta) = \sqrt{2} \sum_{n=1,3}\Big(\phi_\text{nc}\cos(nmt) + \phi_\text{ns}\sin(nmt)\Big) \;,
\end{equation}
in which $m\ne m_a$ generally because the anharmonicity induced by the self-interaction shifts frequencies relative to the harmonic motion.
Furthermore, the third harmonic is suppressed by a factor of $\lambda$ relative to the fundamental one.

Substituting Eq.~(\ref{eq:phiansatz0}) into Eq.~(\ref{eq:KGback}), retaining contributions up to linear in $\lambda$ and discarding terms of
order $H/m_a$ (which decays rapidly below unity after the onset of the axion oscillations), we arrive at the following set of algebraic
(rather than differential) equations:
\begin{align}
  \big(m_a^2-m^2\big) a^2 \phi_\text{1c} &= \frac{a^2\lambda}{4} \big(\phi_\text{1c}^3+\phi_\text{1c}\phi_\text{1s}^2\big) \\
  \big(m_a^2-m^2\big) a^2 \phi_\text{1s} &= \frac{a^2\lambda}{4} \big(\phi_\text{1s}^3+\phi_\text{1s}\phi_\text{1c}^2\big) \nonumber
\end{align}
for the fundamental mode, and
\begin{align}
  \big(m_a^2-9m^2\big) a^2 \phi_\text{3c} &= \frac{a^2\lambda}{12} \big(\phi_\text{1c}^3-3\phi_\text{1c}\phi_\text{1s}^2\big) \\
  \big(m_a^2-9m^2\big) a^2 \phi_\text{3s} &= \frac{a^2\lambda}{12} \big(3\phi_\text{1c}^2\phi_\text{1s}-\phi_\text{1s}^3\big) \nonumber
\end{align}
for the third harmonic. Hence, the relative frequency shift is given by
\begin{equation}
  \label{eq:FreqAnharmonic}
  \Delta m_a^2 \equiv \frac{m^2-m_a^2}{m_a^2} = -\frac{1}{4 f^2}\big(\phi_\text{1c}^2+\phi_\text{1s}^2\big) \;,
\end{equation}
while the amplitude of the third harmonic reads
\begin{align}
  \label{eq:AmpAnharmonic}
  \phi_\text{3c} &= -\frac{1}{96 f^2}\big(\phi_\text{1c}^3-3\phi_\text{1c}\phi_\text{1s}^2\big) \\
  \phi_\text{3s} &= \frac{1}{96 f^2}\big(\phi_\text{1s}^3-3\phi_\text{1s}\phi_\text{1c}^2\big) \nonumber \;.
\end{align}
The scaling $\phi_\text{1c},\phi_\text{1s}\sim a^{-3/2}$ implies that the frequency shifts decays as $a^{-3}$, whereas the amplitude of the third
harmonic behaves like $\phi_\text{3c},\phi_\text{3s}\sim a^{-9/2}$, that is, it decays quickly with the expansion of the Universe.
Furthermore, both the squared frequency shift $\Delta m_a^2$ and the ratio $|\phi_3|/|\phi_1|$ have a present-day amplitude
$\phi_1^2/m_a^2\sim \rho_a/(m_a f)^2\sim 10^{-18}$. They become of order unity only when $a$ approaches $a_\text{osc}$ (signaling the breakdown
of our perturbative approach).
Notwithstanding, we will see in the forthcoming Section that the third and, more generally, higher order harmonics should be taken into account 
because they generate contributions to the self-interaction potential at all order in the coupling $\lambda$.

\section{Coherent oscillations in the presence of perturbations}
\label{sec:perturbations}

In this Section, we investigate how the axion coherent oscillations affect the equation of motion of the slowly-varying envelop of the axion
condensate when perturbations around the background are included.
We focus on the backreaction of the time-dependent gravitational potential, and on the third harmonic induced by the quartic coupling.

\subsection{Time dependence of the gravitational potential}
\label{sec:gravosc}

Let us explore first how oscillations in the axion density and pressure backreact on the gravitational potential $\Psi$.

To carry out this calculation, we start from the (scalar) Einstein equations 
\begin{align}
  -3{\cal H}^2 \Psi - 3{\cal H}\dot{\Psi} - \partial_i\partial^i\Psi &= 4\pi G a^2 \delta \rho_a \\
  \left(2\frac{\ddot{a}}{a}-{\cal H}^2\right)\Psi + 3 {\cal H}\dot{\Psi} + \ddot{\Psi} &= 4\pi G a^2 \delta P_a \nonumber \;,
\end{align}
in the longitudinal gauge, in which we set $\Psi=\Phi$ (this is justified so long as there is no significant anisotropic stress).
The density and pressure perturbations are given by $\delta\rho_a=\rho_a-\bar\rho_a$ and $\delta P_a = P_a - \bar{P}_a$, in which the
background values $\bar\rho_a$ and $\bar{P}_a$ are evaluated as in Eqs.(\ref{eq:barrho}) and (\ref{eq:barP}).
As first recognized by \cite{Khmelnitsky/Rubakov:2013}, oscillations of the background pressure on a timescale $(2m_a)^{-1}$, see Eq.(\ref{eq:barP}),
leave a similar imprint on the gravitational potential. On decomposing the latter as
\begin{align}
  \label{eq:oscigrav}
  \Psi &= \Psi_0 + \Psi_\text{2c}\cos(2 m_a t) + \Psi_\text{2s} \sin(2 m_a t) \\
  &= \Psi_0 + \Psi_2\, e^{-2im_a t} + \Psi_2^*\, e^{2im_a t} \nonumber \;,
\end{align}
we can solve for the slow- ($\Psi_0$) and fast-oscillating ($\Psi_\text{2c}$ and $\Psi_\text{2s}$) contributions to the potential. Separating out
the different harmonics and taking into account the dominant terms solely, we obtain
\begin{align}
  \Delta_\vx \Psi_0 &= 4 \pi G a^2 \big(\rho_0-\bar\rho_0\big) \label{eq:Poisson} \\
  -m_a^2 \Psi_\text{2s} &= \pi G \big(P_\text{2s}-\bar P_\text{2s}\big) \nonumber \\
  -m_a^2 \Psi_\text{2c} &= \pi G \big(P_\text{2c}-\bar P_\text{2c}\big) \nonumber \;.
\end{align}
The last two relations follow from the Einstein equation for the isotropic stress.
The average pressures $\bar P_\text{2s}$ and $P_\text{2c}$ can generally be expressed as
\begin{align}
  \bar P_\text{2s} &= -\bar\rho_0\sin(2\vartheta) \\
  \bar P_\text{2c} &= -\bar\rho_0\cos(2\vartheta) \nonumber \;,
\end{align}
where, again, $\vartheta=3\pi/8$ in the simplified model discussed in Appendix \S\ref{app:homKG}.

These equations show that the amplitude of $\Psi_\text{2s}$ and $\Psi_\text{2c}$ is 
\begin{equation}
  \label{eq:estimatePsi2}
  \big\lvert\Psi_\text{2s}\big\lvert \sim \big\lvert\Psi_\text{2c}\big\lvert \sim \frac{k^2}{m_a^2 a^2} \big\lvert\Psi_0\big\lvert \;.
\end{equation}
Unlike the gravitational slip $\Phi-\Psi$ which is independent of scale, the suppression of the fast-oscillating potentials strongly depends on
wavenumber. These become of order $\Psi_0$ on the Compton scale $k\sim m_a a$, which is $\sim 1\hkpc$ at recombination.
        
\subsection{Including a quartic interaction}

We are now in a position to derive the non-relativistic limit of the KG equation, Eq.(\ref{eq:KGsimplified}), taking into account the feedback from
both the oscillations in the gravitational potential and the quartic interaction. 
When the latter is included, one should consider the ansatz 
\begin{equation}
  \label{eq:phiansatz1}
  \phi(\eta,\vx) = \sqrt{2}\,{\rm Re}\!\Big[\psi_1(\eta,\vx)\, e^{-im_at} + \psi_3(\eta,\vx) e^{-3im_a t}\Big]
\end{equation}
as emphasized in the homogeneous case, see Sec.~\S\ref{sec:homogeneous}. 

Because the leading order term $m_a^2a^2\psi_1$ cancels out in the $e^{\pm im_a t}$ harmonic, we shall retain terms up to order $H/m_a$ for $\psi_1$,
which are of the form $\dot{\psi_1}$ and ${\cal H}\psi_1$. By contrast, the mass term $m_a^2 a^2\psi_3$ does not vanish from the $e^{\pm 3im_a t}$ harmonics.
Therefore, the time-dependence of $\psi_3$ can be ignored when $H/m_a\ll 1$. We expect that the higher order harmonics neglected here can also be treated
in the steady-state approximation.
Regarding the gravitational potential, we retains term up to order $(k/m_a a)^2$ times the leading contribution $\Psi_0\psi_1$ such as to include the
feedback from oscillations in the gravitational potentials. 
The harmonics $e^{-im_a t}$ and $e^{-3im_a t}$ yield the equations 
\begin{align}
  \label{eq:GP1}
  i a\bigg(\partial_\eta\psi_1+\frac{3}{2}\mathcal{H}\psi_1\bigg)
  &= -\frac{1}{2m_a} \partial_i\partial^i\psi_1 \\ & \quad + m_a a^2\bigg[\left(\Psi_0 - \frac{1}{8f^2}|\psi_1|^2\right)\psi_1 \nonumber \\
    &\qquad -3\Psi_2\psi_1^*-\frac{1}{8f^2}(\psi_1^*)^2\psi_3\bigg] \nonumber \\
  \partial_i\partial^i\psi_3  +m_a^2 a^2 &\bigg(8 \psi_3 +\frac{1}{12 f^2}\psi_1^3\bigg)=0
  \label{eq:GP3}\;.
\end{align}
Owing to the third harmonic, the fast oscillatory terms usually neglected in the original Gross-Pitaevskii equation have now vanished from Eq.(\ref{eq:GP1})
\cite[see also][]{Namjoo:2017nia}.
Note also that, 
while the second equation is linear in the dimensionless coupling $\lambda$, the first includes terms up to order ${\cal O}(\lambda^2)$ to emphasize that the
backreaction of $\psi_3$ on $\psi_1$ is only second-order.
The harmonics $e^{im_a t}$ and $e^{3im_a t}$ furnish equations for $\psi_1^*$ and $\psi_3^*$, which are precisely the complex conjugate of Eqs.(\ref{eq:GP1}) --
(\ref{eq:GP3}). In particular, the complex potential $\Psi_2\equiv \frac{1}{2}(\Psi_\text{2c}+i\Psi_\text{2s})$ is replaced by $\Psi_2^*$ in those equations.
Finally, the minus sign in front of $3\Psi_2\psi_1^*$ is, as we shall see shortly, important for the physical interpretation of this term.
        
Following \cite{Desjacques/etal:2018}, we rescale the coordinates and the fields according to
\begin{gather}
  \label{eq:VarTransform}
  \eta \to \frac{1}{m_a}\eta=\tilde \eta\;,\quad \vx=\tilde \vx\;, \quad 
  \psi_i\to \frac{m_a}{f}\psi_i=\tilde \psi_i\;, \\  \rho\to \frac{1}{f^2}\rho = \tilde \rho\;, \quad 
  \Psi_i\to m_a^2 \Psi_i=\tilde \Psi_i \nonumber \;.
\end{gather}
Note that this is not the only possible coordinate redefinition which removes (at least part of) the characteristic scales of the problem
\cite[see, for instance,][for a different choice]{Levkov:2016rkk}. The prescription Eq.(\ref{eq:VarTransform}) has the advantage that corrections to the
dimensionless Gross-Pitaevskii-Poisson system arise with factors of $1/m_a^2$.

Upon applying this coordinate and field transformation to Eqs.~(\ref{eq:GP1}) and (\ref{eq:GP3}), these equations can be recast into the form 
\begin{align}
  i a\bigg(\partial_\eta\psi_1+\frac{3}{2}\mathcal{H}\psi_1\bigg)
  &= -\frac{1}{2} \partial_i\partial^i\psi_1+ a^2\bigg[\left(\Psi_0 - \frac{1}{8}|\psi_1|^2\right)\psi_1 \nonumber \\
    & \qquad -3\Psi_2\psi_1^*-\frac{1}{8}(\psi_1^*)^2\psi_3\bigg]
  \label{eq:GP1s} \\
  \partial_i\partial^i\psi_3  &+a^2 \bigg(8 m_a^2 \psi_3 +\frac{1}{12}\psi_1^3\bigg) =0
  \label{eq:GP3s}
\end{align}
We omitted the tildes of the new coordinates and fields to avoid clutter. 

These equations must be supplemented by ``Poisson'' equations for the gravitational potentials $\Psi_0$, $\Psi_\text{2c}$ and $\Psi_\text{2s}$.
Assuming that both the energy density and pressure are dominated by the mass term $\sim m_a^2|\psi_1|^2$ (so that $\psi_1$ can be interpreted as a wave function),
these equations read
\begin{align}
  \Delta_\vx \Psi_0 &= 4 \pi \tilde G a^2 \big(\rho_0-\bar\rho_0\big) \label{eq:Poisson} \\
  -m_a^2 \Psi_\text{2s} &= \pi \tilde G \big(P_\text{2s}-\bar P_\text{2s}\big) \nonumber \\
  -m_a^2 \Psi_\text{2c} &= \pi \tilde G \big(P_\text{2c}-\bar P_\text{2c}\big) \nonumber
\end{align}
in the transformed coordinates, where the gravitational constant $\tilde G=(m_a f)^2/m_P^2$ now has dimension of energy squared.
Furthermore, the general expression for the energy and pressure of a scalar field yields
\begin{align}
  \rho_0 &= |\psi_1|^2 \\
  P_\text{2s} &= \frac{i}{2} \Big[\psi_1^2-(\psi_1^*)^2\Big] \nonumber \\
  P_\text{2c} &= -\frac{1}{2} \Big[\psi_1^2+(\psi_1^*)^2\Big] \nonumber \;.
\end{align}
Finally, the background density and pressure must be rescaled according to $\bar\rho_0\to \frac{1}{f^2}\bar\rho_0$, $\bar P_i\to \frac{1}{f^2}\bar P_i$.

Note that it is not possible to absorb, through any coordinate and field redefinition, the factor of $m_a^2$ which appears in the equation for $\psi_3$
and for the oscillatory pieces $\Psi_\text{2s}$ and $\Psi_\text{2c}$ of the gravitational potential.
In other words, the relative importance of the higher harmonics and the oscillatory potential is an absolute scale that cannot be transformed away.

\section{Discussion}
\label{sec:discussion}

We will now discuss the implications of the system Eqs.~(\ref{eq:GP1s}) -- (\ref{eq:Poisson}).
While we focus on the evolution of the fundamental mode $\psi_1$,
we emphasize that the considerations drawn in this Section also apply to the higher-order harmonics.

\subsection{Backreaction of the higher harmonics}

As emphasized above, $\psi_3$ is evaluated in steady-state approximation because, unlike $\psi_1$, the mass term does not vanish.
When the kinetic term $\partial_i\partial^i\psi_3$ can also be neglected
~\footnote{This is similar to the Thomas-Fermi approximation \cite[see][in the context of Bose-Einstein Condensates]{Dalfovo:1999zz} which, however, is
  applied to the fluid representation of the GP equation, cf. \cite{Boehmer:2007um}.}, which is
true so long as $k\lesssim m_a a$, Eq.(\ref{eq:GP3s}) furnishes a simple algebraic relation between the fundamental mode and the third harmonic,
\begin{equation}\label{eq:psi1 psi3 relation}
  \psi_3 = -\frac{1}{96m_a^2}\psi_1^3 \;.
\end{equation}
Once this solution is substituted into the equation for the fundamental mode, it changes the content of the square brackets in Eq.(\ref{eq:GP1s}) into 
\begin{equation}
  \Psi_0\psi_1 - \frac{1}{8}|\psi_1|^2\left(1-\frac{1}{96 m_a^2}|\psi_1|^2\right)\psi_1 -3 \Psi_2\psi_1^* \;.
\end{equation}
This shows that, at order $\lambda^2$, the third harmonic ``renormalizes'' the self-interaction potential experienced by the fundamental mode into
\begin{equation}
  -\frac{1}{8f^2}|\psi_1|^2\psi_1 \to -\frac{1}{8f^2}|\psi_1|^2\psi_1 + \frac{1}{768 f^4}|\psi_1|^4 \psi_1 \;,
\end{equation}
where we have momentarily switched back to dimensionfull coordinates and fields.

We expect that higher-order harmonics will introduce corrections at all orders $f^{-2n}$, as discussed in Appendix \S\ref{app:higherharmonics} .
In other words, the effective self-interaction felt by the fundamental mode $\psi_1$ is not the quartic interaction of the ``bare'' Lagrangian.
Therefore, the heuristic procedure advocated by \cite{Eby:2016cnq}, in which $\phi^{2n}$ in the bare Lagrangian is replaced by $|\psi|^{2n}$ in the GP equation,
must be applied with care.
Although these higher-order contributions to the self-interaction are suppressed by powers of $\rho_a/(m_af)^2$, these may have implications for the stability
of equilibrium configurations as investigated in, e.g., \cite{Ruffini:1969qy,vv,Boehmer:2007um,Chavanis:2011zi,Desjacques/etal:2018}.
In these studies, one either solves a differential reflecting the hydrostatic equilibrium, or attempts to minimize the energy functional using an educated guess
for the solution. In all cases, the stability sensitively depends on the behavior at short distances or, equivalently, at high densities where corrections to
the self-interaction become important.
Obviously, higher-order bare interactions, if present, will become relevant at high densities and generate additional corrections proportional to $f^{-2n}$
\cite{Eby:2016cnq,Schiappacasse:2017ham,Chavanis:2017loo,visinelli/etal:2018}.

When the kinetic term in Eq.(\ref{eq:GP3s}) cannot be neglected, the latter takes the form of the inhomogeneous Helmholtz (elliptic) equation.
Namely, on defining
\begin{equation}
  \theta=8m_a^2 a^2\psi_3\;, \qquad \boldsymbol{\xi}=\sqrt{8}m_aa\vx \;,
\end{equation}
we can write Eq.(\ref{eq:GP3s}) as 
\begin{align}
  \Delta_{\boldsymbol{\xi}}\theta+\theta= - \frac{a^2}{12}\psi_1^3 \;.
\end{align}
This equation can be solved using Green's functions. For the real part of $\theta(\boldsymbol{\xi})$ for instance, with $\tilde\theta\equiv {\rm Re}(\theta)$,
the general solution is
\begin{equation}
  \label{eq:greenpsi3}
  \tilde\theta(\boldsymbol{\xi}) = \tilde\theta_0(\boldsymbol{\xi})-\frac{a^2}{12}\int\!d^3\xi_0\,G(\boldsymbol{\xi},\boldsymbol{\xi}_0)\,
              {\rm Re}\big[\psi_1^3(\boldsymbol{\xi}_0)\big]\;,
\end{equation}
where $\tilde\theta_0$ is the solution to the homogeneous Helmholtz equation, $( \Delta_{\boldsymbol{\xi}}+1)\tilde\theta_0(\boldsymbol{\xi})=0$, and the stationary
wave Green's function
\begin{equation}
G(\boldsymbol{\xi},\boldsymbol{\xi}_0) = - \frac{\cos(|\boldsymbol{\xi}-\boldsymbol{\xi}_0|)}{4\pi|\boldsymbol{\xi}-\boldsymbol{\xi}_0|}
\end{equation}
is suitable for the description of the steady-state solutions we are interested in.
In the case of spherical symmetry, the homogeneous Helmholtz equation reduces to the Lane-Emden equation with $n=1$.

In practice, in a numerical cosmological simulation of the large scale structure, the coupled $(\psi_1,\psi_3)$ system could be solved using a relaxation approach,
in which the solution to Eq.(\ref{eq:GP1s}) with $\psi_3\equiv 0$ provides a good initial guess for $\psi_1$.
This trial solution is then substituted into Eq.(\ref{eq:greenpsi3}) assuming the homogeneous solution vanishes, $\tilde\theta_0\equiv 0$
(This follows from the fact that $\psi_3$ is sourced only by a non-vanishing $\psi_1$), etc., until the required convergence is achieved.
Note that the solution will differ from the simple scaling $\psi_3\propto \psi_1^3$, so that the product $(\psi_1^*)^2\psi_3$ takes the general form of a
complex function times $\psi_1$. This brings us to the backreaction of the gravitational potential and the presence of dissipative terms.

\subsection{Backreaction of the gravitational potential}

While, in Eq.(\ref{eq:GP1s}), the purely imaginary coefficient $ia{\cal H}$ encodes the dilution of the (conserved) number of particles owing to the expansion, and
purely real terms such as $a^2 \Psi_0$ or $-\frac{a^2}{8}|\psi_1|^2$ correspond to the leading gravitational and self-interaction contributions to the energy, the
term $\Psi_2\psi_1^*$ implies a generally complex coefficient $\Psi_2\psi_1^*/\psi_1$ and, as such, could provide a source of dissipation.

In order to further investigate this issue, we write $3\Psi_2\psi_1^*$ in Eq.(\ref{eq:GP1s}) as
\begin{align}
  3\Psi_2\psi_1^* &= \alpha\Big[|\psi_1|^2\psi_1 + \big(\bar P_\text{2c}+i\bar P_\text{2s}\big)\psi_1^*\Big] \\
  &= \alpha\Big[|\psi_1|^2\psi_1 - \bar\rho_0 e^{2i\vartheta} \psi_1^*\Big] \nonumber \\
  &= \alpha\Big[|\psi_1|^2\psi_1 - \bar\rho_0 e^{2i(\vartheta-\varphi)} \psi_1\Big] \nonumber \;.
\end{align}
Here, $\alpha\equiv 3\pi \tilde G/2m_a^2$ is a real positive constant,
and $\varphi(\eta,\vx)$ is the argument or phase of $\psi_1(\eta,\vx)$, i.e. $\varphi={\rm arg}(\psi_1)$. 
In the last equality, the amplitude of first term relative to the dominant contribution $\Psi_0\psi_1$ is $(k/m_a a)^2$.
Therefore, it leads to a significant correction to the usual Newtonian self-interaction $\Psi_0\psi_1$ on scales $k\gtrsim m_a a$ smaller than the Compton length.
Furthermore, since it can be written as a real coefficient multiplying $\psi_1$, it contributes to the momentum conservation equation solely once
the GP system is expressed as fluid equations.
Conversely, the second term cannot generally be recast in the form of a real coefficient multiplying $\psi_1$ and, thus, contributes to both the
continuity and momentum conservation equations (see below).
Finally, since the coupling $\Psi_2\psi_1^*$ is absent at the homogeneous level, the phase $\la\varphi\ra$ of the homogeneous solution $\psi_1$ must satisfy
$\la\varphi\ra\equiv\vartheta$. This is no longer the case when inhomogeneities are present.

The Madelung transformation $\psi_1(\eta,\vx) \equiv \sqrt{\rho(\eta,\vx)} e^{i\varphi(\eta,\vx)}$ leads to a reformulation of Eq.~(\ref{eq:GP1s}) as
(real) hydrodynamical equations \cite{Madelung:1927,Bohm:1952}.
The imaginary part of $\Psi_2\psi_1^*$, proportional to $\bar\rho_0\sin\!\big(2(\vartheta-\varphi)\big)$, manifests itself as a source term in the
continuity equation, i.e.
\begin{equation}
\partial_\eta\rho+3{\cal H}\rho+\grad_\vx\big(\rho \vu\big) = 2\alpha a\rho\bar\rho_0\sin\!\big(2(\la\varphi\ra-\varphi)\big) \;.
\end{equation}
Such a term arises because the Lagrangian leading to the equation of motion Eqs.(\ref{eq:GP1s}) is no longer invariant under a global phase
transformation $\varphi(\eta,\vx)\to\varphi(\eta,\vx) + \varphi_0$ owing to the presence of a term $\Psi_2(\psi_1^*)^2+\Psi_2^*\psi_1^2$.
As a result, the total probability $\int\! d^3x\,|\psi_1|^2$ is not
conserved~\footnote{We have not considered the possibility of a time- and position-dependent chemical potential $\mu(\eta,\vx)$}.
Notwithstanding, the amplitude of this source term relative to $\dot{\rho}\sim {\cal H}\rho$ decreases like $a^{-3/2}$ and does not depend on the axion
decay constant $f$. A rough estimate yields
\begin{equation}
  \frac{{\rm source}}{\dot{\rho}} \sim 10^{-13} a^{-3/2} \left(\frac{m_a}{10^{-22}\eV}\right)^{-1} \;,
\end{equation}
where ``source'' designates the term proportional to $\sin\!\big(2(\la\varphi\ra-\varphi)\big)$
and the scaling $a^{-3/2}$ is valid in matter domination. 
Hence, for our fiducial axion mass, the source term is negligible throughout the whole period of coherent oscillations, so that $|\psi_1|^2$ can still be
interpreted as a probability density.

The real part of $\Psi_2\psi_1^*$ adds a new contribution to the Hamilton-Jacobi equation,
\begin{align}
  \label{eq:HamiltonJ}
  a^{-1} \partial_\eta\varphi &+ \frac{1}{2}u^2 + Q + h+ \Psi_0 \\
  & \quad - \alpha\Big[\rho-\bar\rho_0 \cos\!\big(2(\la\varphi\ra-\varphi)\big)\Big] =0 \nonumber \;.
\end{align}
Here, $\varphi$ emulates the classical action. Therefore, $\la\varphi\ra\equiv\vartheta$ is the value of the classical action for the homogeneous solution.
Furthermore, $u$ is the modulus of the (curl-free) peculiar velocity field,
\begin{equation}
\vu\equiv a^{-1}\grad_\vx\varphi \;,
\end{equation}
we have used the relation Eq.~(\ref{eq:psi1 psi3 relation}) to compute the enthalpy per unit mass as
\begin{equation}
h(\rho) = -\frac{\rho}{8}\left(1-\frac{\rho}{192 m_a^2}\right) \;,
\end{equation}
and
\begin{equation}
  Q=-\frac{1}{2a^2}\frac{\Delta_\vx\sqrt{\rho}}{\sqrt{\rho}}
\end{equation}
is the so-called ``quantum'' potential, which can be interpreted as an effective pressure arising from the large Compton wavelength of the particles.
In light of the work of \cite{BialynickiBirula:1976zp,Chavanis:2011jx,Chavanis:2017bwo}, the first term $-\alpha\rho$ of the square brackets
can be interpreted as a potential with negative effective temperature or, equivalently, negative pressure.
This reflects the fact that the oscillatory piece of the gravitational potential tends to confine the particles and, thus, produces an attractive force
which counteracts the effect of the ``quantum'' pressure.

To understand the physical meaning of the second term, we note that, at leading order, it gives a contribution
\begin{equation}
-2\alpha\bar\rho_0\big(\varphi-\la\varphi\ra\big)^2
\end{equation}
to the Hamilton-Jacobi equation. Taking the gradient of Eq.~(\ref{eq:HamiltonJ}), we retrieve the momentum conservation equation
\begin{align}
  \label{eq:euler}
  \frac{d\vu}{d\eta} + {\cal H}\vu &= -\grad_\vx\Big( Q +h+ \Psi_0-\alpha\rho\Big)  \\
  & \qquad +4\alpha a \bar\rho_0\big(\varphi-\la\varphi\ra\big)\vu \nonumber \;.
\end{align}
It is evident that the term involving $\varphi-\la\varphi\ra$ plays the role of a friction when $\varphi<\la\varphi\ra$, while the opposite (energy injection)
is true when $\varphi>\la\varphi\ra$. Since the linearized continuity equation gives $\delta\sim -\grad_\vx\cdot\vu\sim -\Delta_\vx\varphi $, we expect $\varphi$
to be convex (concave) in overdense (underdense) regions. Therefore, friction preferentially occurs in underdense regions.
This conclusion remains valid when the full cosine is taken into account. 
Eq.(\ref{eq:euler}) shows that the amplitude of $\alpha\grad_\vx\rho$ relative to the usual Newtonian force is $(k/m_a a)^2$ independently of
time, in agreement with our estimate Eq.(\ref{eq:estimatePsi2}).
By contrast, the ``friction'' term proportional to $(\varphi-\la\varphi\ra)\vu$ is suppressed by a factor $H/m_a$.
Consequently, it is significant only at the onset of axion coherent oscillations.

\section{Conclusions}
\label{sec:conclusions}

We have explored the backreaction of coherent oscillations on the equation of motion of the axion condensate beyond the usual approximations made in the derivation
of the Gross-Pitaevskii (GP) equation.
Although we specialized our results to axions with mass $10^{-22}\eV$, our findings are relevant to any particle in the form of a Bose-Einstein condensate.

Axion self-interactions require that one takes into account harmonics beyond the fundamental mode.
Ref.~\cite{visinelli/etal:2018} recently emphasized that higher harmonics must be considered when the axion field becomes sensitive to the full interaction potential
as is the case of ``dense'' axion stars.
Here, we have pointed out that, even in the case of a simple Lagrangian bare quartic coupling $\lambda\phi^4$, higher-oder harmonics backreact on the evolution of the
fundamental mode and ``renormalizes'' the self-interaction by adding contributions at all orders beyond the quartic term.
Our findings may have implications for global stability analyzes of axion condensates, in which the stability region is inferred by minimizing the energy functional 
of trial equilibrium solutions \cite[see, e.g.,][]{Chavanis:2011uv}.

We have also shown that the rapidly oscillating gravitational potential produced by the axion coherent oscillations \cite[see, e.g.,][]{Khmelnitsky/Rubakov:2013}
backreacts on the equation of motion of the axion condensate mainly in the form of an additional attractive force.
This is made explicit through a reformulation of the GP system as (real) hydrodynamical equations.
This force, which is suppressed by a factor of $(k/m_a a)^2$ relative to the usual Newtonian term, confines the particles and, thus, opposes the effect of the
``quantum'' pressure. Oscillations of the gravitational potential also introduce a ``friction'' into 
the momentum conservation equation, and a source term in the continuity equation. However, these last two effects are suppressed by $H/m_a$ and, thus, are only
significant at the onset of axion oscillations.

To derive these results, we have retained terms that contribute at linear order solely.
Second-order contributions like the anisotropic stress, which are also suppressed by gradients, should become relevant at the Compton length of the particle and,
thus, be taken into account in the hydrodynamic gradient expansion when $k\gtrsim m_a a$.
Clearly, to which extent the evolution of the non-relativistic axion condensate follows the original GP equation is quite uncertain when $k/a$ becomes comparable to
the Compton length.
Overall, a more rigorous and systematic treatment of the non-relativistic limit of the GP system, along the lines of \cite{Namjoo:2017nia,Eby:2017teq} for instance,
would be desirable in order to better control any perturbative calculation. 

Fluctuations in the Lyman-$\alpha$ forest observed in the spectra of distant quasars set tight constraints on the viable mass $m_a$ of ultra-light axions when the
latter accounts for all the dark matter: an axion mass in the range $10^{-22} - 10^{-21}\eV$ is excluded at 95\% C.L. \cite{Irsic:2017yje,Armengaud:2017nkf}.
Using simulations which include the quantum pressure during the structure formation appears to worsen these limits \cite{Zhang:2017chj}.
Could any of the effects considered here have any impact on the evolution of mildly nonlinear density fluctuations traced by the Lyman-$\alpha$ forest and, more
generally, on structure formation ?
A self-interaction with a decay constant as low as $f\sim 10^{15}\GeV$ does not have a large impact on the low density absorbers responsible for the Lyman-$\alpha$
forest \cite{Desjacques/etal:2018}, but certainly affects the dense, virialized regions of the Universe.
Although $m_a$ is severely constrained, it will be instructive to assess whether the backreaction from the gravitational potential could have any significant effect
on cosmic structure formation. We leave a more detailed analysis to future work.
      	
\acknowledgments

V.D. would like to thank Kfir Blum, Alex Kehagias and, especially, Antonio Riotto for useful discussions on the Gross-Pitaevskii equation.
This work is supported by the Israel Science Foundation (grant no. 1395/16). 

\appendix

\section{Solution to the homogeneous KG equation}
\label{app:homKG}

In this Appendix, we outline how the time evolution of the homogeneous slowly-varying piece $\phi_\text{1c}(\eta)$ and $\phi_\text{1s}(\eta)$ defined
in Eq.(\ref{eq:ansatz1}) can be solved approximately under the assumption that the Universe is filled with axions of density parameter
$\Omega_a\sim 1$ and with radiation.
For viable cosmological models, this must be done numerically until the axion field is deep in the regime of coherent oscillation to ensure accuracy
\cite[see, for instance,][]{Hlozek/Grin/etal:2015}. 
       
\subsubsection{Radiation domination}

Since the axion field is initially at rest in its potential, the initial conditions are
\begin{equation}
  \label{eq:ICs}
  \bar\phi(0) = \sqrt{2}\phi_i\;,\qquad \dot{\bar\phi}(0) = 0 \;.
\end{equation}       
The initial field value $\phi_i$ can be related the present-day axion and radiation energy density $\Omega_a$ and $\Omega_r$ through the requirement
that axion oscillations start when $H=\alpha\, m_a$, where $\alpha$ is an order-unity constant to be specified.
Assuming that the axion energy density at the onset of oscillations is given by $\bar\rho_a(a_\text{osc})\approx m_a^2 \phi_i^2$, we find
\begin{equation}
  \frac{\phi_i}{m_P} = \alpha^{3/4}\sqrt{\frac{3}{8\pi}}\, \frac{\Omega_a^{1/2}}{\Omega_r^{3/8}}\, \left(\frac{H_0}{m_a}\right)^{1/4} \;.
\end{equation}
In the radiation era, the homogeneous KG equation admits the solution \cite[e.g.][]{Marsh:2017hbv}
\begin{equation}
  \label{eq:solrad}
  \bar\phi(\eta) = c_1 \frac{J_{1/4}(m_at)}{t^{1/4}}+c_2 \frac{Y_{1/4}(m_at)}{t^{1/4}}
\end{equation}
The initial conditions (\ref{eq:ICs}) require $c_2=0$, which leads to the solution
\begin{equation}
  \label{eq:KGhomsolrad}
  \bar\phi(\eta) = \left(\frac{8}{m_a}\right)^{1/4} \Gamma\!\left(\frac{5}{4}\right)\, \phi_i\, \frac{J_{1/4}(m_at)}{t^{1/4}}\;,
\end{equation}
where $H_0$, $\Omega_a$ and $\Omega_r$ are the present-day Hubble rate, axion and radiation energy density.
In the time window $\eta_\text{osc}\lsim \eta\lsim \eta_\text{eq}$, which is always significant for an axion mass around the fiducial value
$m_a=10^{-22}\eV$ (for which $a_\text{osc}\sim 10^{-7}-10^{-6}$), we can use the asymptotic expansion of the Bessel function $J_{1/4}(x)$,
\begin{align}
  J_{1/4}(x) &\sim \sqrt{\frac{2}{\pi x}}\bigg[\cos\!\left(x-\frac{3\pi}{8}\right) \\
    &\qquad +\frac{3}{32 x}\sin\!\left(x-\frac{3\pi}{8}\right) + {\cal O}\big(x^{-2}\big) \bigg] \nonumber \;,
\end{align}
to approximate the solution as
\begin{equation}
  \label{eq:approx1}
  \bar\phi(\eta) \sim \frac{2^{5/4}}{m_a^{3/4}}\frac{\Gamma(\frac{5}{4})}{\sqrt{\pi}}\,\phi_i\, \frac{\cos\left(m_at-\frac{3\pi}{8}\right)}{t^{3/4}}
\end{equation}
This result is consistent with the findings of \cite{Noh/Hwang/Park:2017}. 

Note that higher order terms in the asymptotic expansions are suppressed by powers of $1/x\sim H/m_a$.
We now choose the coordinates such that, for $0\leq \eta\leq \eta_\text{eq}$, the cosmic time is given by
\begin{equation}
  t = \frac{1}{2} a_0^2 H_0\sqrt{\Omega_r}\eta^2 = \frac{1}{2 H_0\sqrt{\Omega_r}}\left(\frac{a}{a_0}\right)^2 \;,
\end{equation}
where $a_0$ is the present-day value of the scale factor.
On substituting this relation into Eq.(\ref{eq:approx1}) and comparing with Eq.(\ref{eq:ansatz1}), we arrive at
\begin{align}
  \label{eq:phi_rad}
  \frac{\phi_\text{1c}^\text{RD}(\eta)}{m_P} &= \alpha^{3/4}
  \frac{\sqrt{3\Omega_a}}{\pi}\Gamma\!\left(\frac{5}{4}\right)\left(\frac{H_0}{m_a}\right) \cos\!\left(\frac{3\pi}{8}\right) \bigg(\frac{a_0}{a}\bigg)^{3/2} \\
  \frac{\phi_\text{1s}^\text{RD}(\eta)}{m_P} &= \alpha^{3/4}
  \frac{\sqrt{3\Omega_a}}{\pi}\Gamma\!\left(\frac{5}{4}\right)\left(\frac{H_0}{m_a}\right) \sin\!\left(\frac{3\pi}{8}\right) \bigg(\frac{a_0}{a}\bigg)^{3/2}
  \nonumber \;,
\end{align}
which turns out to be also valid in matter domination. This shows that the phase shift is $\vartheta\equiv 3\pi/8$.
In Sec.~\S\ref{sec:discussion}, we demonstrate that $\vartheta$ is equal to the phase $\la\varphi\ra$ of the homogeneous solution for the fundamental
mode $\psi_1$.

\subsubsection{Matter domination}

To see this, we start from the general solution to Eq.(\ref{eq:KGhom}) in matter domination $a>a_\text{eq}$, which is
\begin{equation}
  \label{eq:solmat}
  \bar\phi(\eta) = c_1 j_0(m_a t) + c_2 y_0(m_a t) \;,
\end{equation}
where $j_0(x)$ and $y_0(x)$ are spherical Bessel functions. Substituting 
\begin{equation}
  j_0(x) =\frac{\sin(x)}{x}\;,\qquad y_0(x) =-\frac{\cos(x)}{x} 
\end{equation}
and matching with Eq.(\ref{eq:ansatz1}), the unknown coefficients $c_1$ and $c_2$ are related to $\phi_\text{1c}^\text{MD}$ and $\phi_\text{1s}^\text{MD}$
through
\begin{align}
  c_1 &= \frac{2\sqrt{2}}{3} \phi_\text{1s}^\text{MD}\, \left(\frac{H_0}{m_a}\right)^{-1}\Omega_a^{-1/2} \left(\frac{a}{a_0}\right)^{3/2} \\
  c_2 &= -\frac{2\sqrt{2}}{3} \phi_\text{1c}^\text{MD}\,\left(\frac{H_0}{m_a}\right)^{-1}\Omega_a^{-1/2} \left(\frac{a}{a_0}\right)^{3/2} \nonumber
\end{align}
upon using the fact that, for $\eta\gtrsim\eta_\text{eq}$, the cosmic time is given by
\begin{equation}
  t \simeq \frac{2}{3H_0\sqrt{\Omega_a}}\left(\frac{a}{a_0}\right)^{3/2} \;.
\end{equation}
A comparison with the asymptotic solution Eq.(\ref{eq:approx1}) in radiation domination immediately leads to
\begin{equation}
  \label{eq:phi_mat}
  \phi_\text{1c}^\text{MD}=\phi_\text{1c}^\text{RD} \;,\qquad \phi_\text{1s}^\text{MD}=\phi_\text{1s}^\text{RD} \;.
\end{equation}
This shows that, when $a_\text{osc}\ll a_\text{eq}$ (so that the RD solution is well approximated by the term proportional to $\cos(m_a t-3\pi/8)$ as $a$
approaches $a_\text{eq}$), the slowly-varying amplitudes $\phi_\text{1c}(t)$ and $\phi_\text{1s}(t)$ are the same in both matter and radiation domination.
This gives the adiabatic approximation to the evolution of the mean axion field $\bar\phi(\eta)$.

The free parameter $\alpha$ can eventually be constrained upon requesting that the present-day axion density parameter be $\Omega_a$.
Substituting $\phi_\text{1s}^\text{MD}$ and $\phi_\text{1c}^\text{MD}$ into the slowly varying part $\bar\rho_0$ of the axion background energy density,
Eq.(\ref{eq:barrho}), we find
\begin{equation}
  \bar\rho_0(a_0) = \alpha^{3/2} \frac{8}{\pi}\Gamma\!\left(\frac{5}{4}\right)^2 \Omega_a \rho_\text{cr,0} \equiv \Omega_a \rho_\text{cr,0}\;,
\end{equation}
which implies $\alpha\approx 0.6$. This result is, of course, only valid within the assumptions of the model considered here. In any case, $\alpha$ will 
always be of order unity. 
        
\section{Derivation of Eqs.(\ref{eq:GP1}) and (\ref{eq:GP3})}
\label{app:derivationGP13}
        
In this Appendix, we present a concise derivation of Eqs.(\ref{eq:GP1}) and (\ref{eq:GP3}).
The difficulty resides in the fact that there are a few distinct physical scales that can be adjusted independently and, therefore, a few small parameters.
Our goal is to identify the next-to-leading-order contributions arising from the self-interaction term and from the oscillatory potential to the standard
GP equation.

We start from Eq.(\ref{eq:KGsimplified}), in which we substitute our ansatz for both the axion field, Eq.(\ref{eq:phiansatz1}), and the oscillatory
gravitational potential, Eq.(\ref{eq:oscigrav}).
Assuming that $\psi_1(\eta,\vx)$ and $\psi_3(\eta,\vx)$ vary slowly on a timescale $H^{-1}$, the time derivatives of the axion field $\phi$ are given by
\begin{align}
\dot \phi&=\frac{1}{\sqrt{2}}\Big[e^{-im_at}\big(\dot \psi_1-im_aa\psi_1\big) +e^{im_at}\big(\dot \psi_1^* +im_aa\psi_1^*\big) \nonumber \\ 
&\quad +e^{-3im_at}\big(\dot \psi_3-3im_aa\psi_3  \big)+e^{3im_at}\big(3im_aa\psi_3^* +\dot \psi_3^* \big)\Big] \nonumber\\
\ddot{ \phi}&=\frac{1}{\sqrt 2}\Big[-e^{-im_at}\big(im_a\dot a  \psi_1 +2im_aa\dot \psi_1 +m_a^2 a^2 \psi_1 \big) \nonumber \\ 
&\quad +e^{im_at}\big(im_a\dot a\psi_1^* +2im_aa\dot\psi_1^* -m_a^2a^2 \psi_1^* \big) \nonumber \\
&\quad -e^{-3im_at}\big(3im_a\dot a  \psi_3 -6im_aa  \psi_3 -9m_a^2a^2  \psi_3\big) \nonumber \\
&\quad +e^{3im_at}\big(3im_a\dot a  \psi_3 +6im_aa  \psi_3 -9m_a^2a^2  \psi_3\big) \Big]
\end{align}
upon neglecting terms of order $(H/m_a)^2$ and higher. 
Next, we insert these expressions into the Klein-Gordon equation, Eq.~(\ref{eq:KGsimplified}).
Further discarding terms of the form $\Psi_0\partial_i\partial^i\psi_1$ (which are second-order in perturbations), the surviving terms multiplying
$e^{-im_at}$ and $e^{-3im_at}$ give:
\begin{widetext}
  \begin{align}
    \underline{e^{-im_at}:} &\quad
    \Big(-im_a\dot a\psi_1-2im_aa \dot{ \psi}_1-m_a^2a^2\psi_1\Big)
    -2im_a a \mathcal H \psi_1
    -4\Big(2 m_a^2 a^2 \Psi_2\psi_1^*-im_a a \dot\Psi_0 \psi_1 + 6 m_a^2 a^2 \Psi_2^*\psi_3\Big) \\
    & \quad -\partial_i \partial^i \psi_1 +m_a^2 a^2 \Big[(1+2\Psi_0)\psi_1+2\Psi_2 \psi^*_1+ 2\Psi^*_2  \psi_3\Big]-
    \frac{a^2\lambda}{2\cdot 3!}\Big[3\psi_3(\psi_1^*)^2 + 3|\psi_1|^2\psi_1 + 6|\psi_3|^2\psi_1\Big] = 0 \nonumber \\
    \underline{e^{-3im_at}:} & \quad
    \Big(-3im_a\dot a  \psi_3 -6im_aa  \dot \psi_3 -9m_a^2a^2  \psi_3\Big)-6i m_a a \mathcal H\psi_3
    -4\Big(-3 im_a a \dot\Psi_0\psi_3-2m_a^2 a^2 \Psi_2 \psi_1\Big) \\
    & \quad -\partial_i\partial^i\psi_3 +m_a^2a^2\Big[(1+2\Psi_0)\psi_3+2\Psi_2\psi_1\Big]
    - \frac{a^2\lambda}{2\cdot 3!}\Big[\psi_1^3 + 6|\psi_1|^2\psi_3 + 3|\psi_3|^2\psi_3\Big]=0 \nonumber \;.
  \end{align}
\end{widetext}
The equations for $e^{im_at}$ and $e^{3im_at}$ yield the complex conjugates of these relations and, therefore, do not encode any additional information. 

In the first equation, the mass term $m_a^2 a^2 \psi_1$ cancels out, and we are led to consider terms involving the gravitational potential $\Psi_0$ and $\Psi_2$.
The purely imaginary contribution $i m_a a \dot \Psi_0\psi_1$ is a small correction to $i m_a \dot a \psi_1$ and can, therefore, be neglected.
The terms proportional to $\Psi_2 \psi_1^*$, which are generally complex, are the leading contribution induced by the oscillatory potential.
We shall keep them, but note that they are suppressed by a factor of $(k/m_a)^2$ relative to the usual Newtonian potential $m_a^2a^2 \Psi_0\psi_1$.
Owing to the presence of $\psi_3\propto\lambda$, the term $\Psi_2^*\psi_3$ is suppressed by an additional factor of $\lambda$ relative to $\Psi_2\psi_1^*$ and  
we shall not retain it here. Finally, in the self-interaction, we retain the dominant term $\lambda |\psi_1|^2 \psi_1$ along with the next-to-leading-order
contribution $\lambda (\psi_1^*)^2\psi_3$.

In the second equation, the mass term does not vanish so that all additional contributions can be neglected except for the kinetic term
$\partial_i\partial^i\psi_3$, and the self-interaction. In the latter case, we only keep the dominant term $\lambda \psi_1^3$. 
This can be clearly seen upon dividing the whole equation by $m_a^2 a^2$ so that, e.g., $\frac{m_a \dot a \psi_3}{m_a^2 a^2} \sim (H/m_a)\psi_3$ etc.
The gravitational coupling $\Psi_2\psi_1$ is, in fact, of the same order as $\Psi_2^*\psi_3$ in the former equation.
Both generate corrections of the form $\Psi_2\rho_a\psi_1^*$ in the equation for the fundamental mode. 

After some further simplifications, we arrive at the desired equations (\ref{eq:GP1}) and (\ref{eq:GP3}). 

\section{High Order Harmonics}
\label{app:higherharmonics}

We extend the ansatz \eqref{eq:phiansatz1} to include higher harmonics,
\begin{align}
  \phi(\eta,\vx)=\sqrt{2}\sum_{n\geq 1,\text{$n$ odd}} \bigg(\psi_n e^{-inm_at}+\psi_n^* e^{inm_at}\bigg) \;,
\end{align}
where $\psi_n(\eta,\vx)$ varies on a timescale $\sim H^{-1}$. Owing to the quartic coupling, we expect that odd harmonics solely are present in this expansion so that
$n$ is an odd integer.

As outlined in Sec.~\S\ref{sec:perturbations} and in the previous Appendix, the mass term does not cancel for the higher harmonics, unlike for the fundamental mode.
Therefore, the time derivatives of the Klein-Gordon equation is dominated by the term $\propto n^2 m_a^2$ arising from $\ddot\phi$.
Ignoring the contributions proportional to the gravitational potential, we obtain
\begin{align}
  -(n^2-1)&a^2m_a^2 \psi_n-\partial_i \partial^i \psi_n =  \\
  &\frac{a^2\lambda}{4}\sum_{j,k}^\infty\bigg(A_{jk}^n\psi_j\psi_k\psi_{n-j-k} + B_{jk}^n \psi^*_j \psi_k \psi_{n+j-k}  \nonumber \\ 
  &\qquad + C_{jk}^n \psi_j^* \psi_k^* \psi_{n+j+k}  + D_{jk}^n \psi_j^* \psi_k^* \psi_{-n+j+k}^*\bigg) \nonumber
\end{align}
for $n>3$, where $A_{jk}^n$, ..., $D_{jk}^n$ are combinatoric coefficients.
Since the interaction mixes different harmonics, solving these equations is not trivial even in the Thomas-Fermi limit, in which the kinetic term can be
discarded. Nevertheless, when the right-hand side includes only the leading-order contribution induced by the self-interaction, they simplify to
\begin{equation}
  -(n^2-1)m_a^2 \psi_n \approx \frac{\lambda}{4} \psi_1^2 \psi_{n-2}
\end{equation}
for $n>3$ when the kinetic term can be neglected.
This shows that the dominant contribution to $\psi_n$ scales like
\begin{equation}
  \psi_n \sim \frac{1}{f^{n-1}} \psi_1^n \;,\quad\mbox{$n$ odd} 
\end{equation}
and, therefore, the effective self-interaction potential felt by the fundamental mode $\psi_1$ receives contributions at all order in $1/f^2$.
               
\bibliographystyle{prsty}
\bibliography{references}
        
\label{lastpage}
        
\end{document}